# Giant Third-Order Nonlinearity Induced by the Quantum Metric Quadrupole in Few-Layer WTe$_2$

Xing-Yu Liu,[1] An-Qi Wang,[1,*] Dong Li,[1] Tong-Yang Zhao,[1] Xin Liao,[1] and Zhi-Min Liao[1,2,†]

[1]*State Key Laboratory for Mesoscopic Physics and Frontiers Science Center for Nano-optoelectronics, School of Physics, Peking University, Beijing 100871, China*
[2]*Hefei National Laboratory, Hefei 230088, China*

The quantum geometric properties of topological materials underpin many exotic physical phenomena and applications. Quantum nonlinearity has emerged as a powerful probe for revealing these properties. The Berry curvature dipole in nonmagnetic materials and the quantum metric dipole in antiferromagnets have been explored by studying the second-order nonlinear Hall effect. Although the quadrupole moment of the quantum geometric tensor is theoretically predicted to induce higher-order quantum nonlinearity, the quantum metric quadrupole remains experimentally unexplored. Here, we report the quantum metric quadrupole induced third-order nonlinear longitudinal electrical response in few-layer WTe$_2$, persisting up to room temperature. Angle-resolved third-harmonic current-voltage characteristics are found consistent with the intrinsic crystal symmetry of WTe$_2$. Through temperature variation and scaling analysis, we identify the quantum metric quadrupole as the physical origin of the observed third-order longitudinal nonlinearity. Additionally, we determine the angle dependence of the quantum metric quadrupole, establishing third-order nonlinearity as an efficient method for revealing the quantum metric structure.

The quantum metric [1] and Berry curvature [2], as components of the quantum geometric tensor, describe the geometric structures of Bloch electronic states. The Berry curvature, as the curvature of Hilbert space, is well known to generate anomalous Hall effects [3,4] and valley Hall effects [5–7]. The integration of the Berry curvature can yield a Chern number, advancing our understanding of topological phases [8,9] and quantum Hall effects [10,11]. The spatial distributions of the Berry curvature described as dipole [12] and multipole [13] moments have been clearly revealed through nonlinear transport [14–22] and optics [23–25]. Correspondingly, the quantum metric, defined as the amplitude distance between two neighbor quantum states, has recently been recognized as potentially having significant observable effects [26]. The integral of the quantum metric is theoretically predicted to yield the superfluid stiffness in flat bands, showing essential properties of correlated many-body states [27–29]. The distribution of the quantum metric is expected to manifest nonlinear transport phenomena, with its dipole contributing to second-order nonlinear Hall effects observed in MnBi$_2$Te$_4$ [30,31] and Mn$_3$Sn/Pt [32] systems. Despite this, studies on the multipole moments of the quantum metric remain limited.

In addition to the second-order nonlinear Hall effect, the third-order nonlinear Hall effect has been investigated to reveal both the Berry curvature quadrupole [13,22] and the electric field-induced Berry curvature dipole [33,34]. Under an applied electric field **E**, a Berry curvature emerges as $\mathbf{\Omega^E} = \nabla_{\mathbf{k}} \times [G(\mathbf{k})\mathbf{E}]$, where $G(\mathbf{k})$ is the Berry connection polarizability tensor [35,36], intimately connected with the quantum metric. In nonmagnetic materials with time-reversal symmetry, the field-induced Berry curvature exhibits a dipolelike pattern with a zero net integral value in the Brillouin zone [34,37], but a nonzero Berry curvature dipole $D^E$. The generated current $\mathbf{j}' \propto \hat{\mathbf{z}} \times \mathbf{E}(\mathbf{D^E} \cdot \mathbf{E}) \propto |\mathbf{E}|^3$ is confined to the in-plane direction perpendicular to **E**, illustrating the third-order nonlinear Hall effect observed in materials like bulk MoTe$_2$ [33], WTe$_2$ [38], and TaIrTe$_4$ [39].

On the other hand, the quantum metric is connected to band energy correction, affecting band velocity, as proposed in the general semiclassical theory [35,36]. The corrected velocity of the *n*th band with an original band energy $\epsilon_n$ is given by

$$\tilde{\boldsymbol{v}}_n = \frac{\partial \epsilon_n}{\hbar \partial \mathbf{k}} - \frac{e^2}{2\hbar} E_a E_b \partial_{\mathbf{k}} G_{ab}, \qquad (1)$$

where $\tilde{\boldsymbol{v}}_n$ is the corrected band velocity, *a* and *b* denote spatial directions, k is the wave vector, *e* is the electron charge, and $\hbar$ is the reduced Planck constant. Therefore, a third-order longitudinal current, related to the quantum metric quadrupole (QMQ), denoted as $\partial_{\mathbf{k}}\partial_{\mathbf{k}}G_{ab}$, is theoretically expected (Supplemental Material, Note 1 [40]). However, the experimental manifestation of the QMQ remains unexplored.

---

*Contact author: anqi0112@pku.edu.cn
†Contact author: liaozm@pku.edu.cn

Here, we report the QMQ-induced third-order nonlinearity in the transport properties of few-layer Weyl semimetal $WTe_2$. Using lock-in measurements and ac current bias, we observe significant longitudinal and transverse third-order harmonic voltages in the devices, persisting up to room temperature. The angle-resolved third-order nonlinearity is found to be consistent with the intrinsic crystal symmetry of $WTe_2$. Through scaling law analysis, we clearly identify contributions from QMQ and extrinsic scatterings. The obtained QMQ-induced nonlinearity is 3 orders of magnitude larger than that in bulk $MoTe_2$ [33] and $TaIrTe_4$ [39].

To provide an intuitive understanding, we first examine the quantum metric and QMQ in a tilted two-dimensional (2D) massive Dirac model, which serves as a simplified framework for various quantum materials [42–44]. As shown in Figs. 1(b) and 1(c), although the components of the quantum metric exhibit monopoles concentrated around the band gap, nonzero QMQs emerge. Because of the anisotropic band structure, $G_{yy}$ exhibits a greater magnitude compared to $G_{xx}$ [Fig. 1(b)], resulting in an increased QMQ along the $y$ direction [Fig. 1(c)]. This QMQ contributes to an additional band velocity, generating third-order currents under an applied electric field, characterized by the longitudinal conductivities $\chi_{xxxx}$ and $\chi_{yyyy}$. Figure 1(d) shows the calculated third-order conductivities, with $\chi_{xxxx}$ and $\chi_{yyyy}$ peaking near the band edges and vanishing within the band gap due to the absence of carriers. Notably, $\chi_{yyyy}$ is about twice $\chi_{xxxx}$ [Fig. 1(c)], corresponding to the larger QMQ and band tilting along the $y$ direction. Furthermore, longitudinal nonlinearity persists even in isotropic bands with time-reversal and inversion symmetries (Supplemental Material, Note 2 [40]), relaxing the symmetry constraints of nonlinear response.

Multilayer $T_d$-$WTe_2$ is noncentrosymmetric with point group $Pm$ (Ref. [14]). The multilayers are stacked with alternating 180° rotations, as shown in Figs. 2(a) and 2(b), with a single mirror line along the high-symmetry $b$ axis (dashed line), perpendicular to the low-symmetry $a$ axis. We studied few-layer $T_d$-$WTe_2$ samples with thicknesses of 6 and 10 layers. As shown in Fig. 2(c), the devices with multiple electrodes arranged in a circular layout were fabricated (see Supplemental Material, Methods [40]). An ac current $I_\omega$ was applied between the source ($S$) and drain ($D$) electrodes, while transverse and longitudinal voltages ($V_\perp$ and $V_\parallel$) were simultaneously measured using lock-in techniques [Fig. 2(d)].

We first conduct measurements at 300 K. As shown in Fig. 2(e), the first-harmonic longitudinal and transverse voltages depend linearly on the magnitude of $I_\omega$. The finite Hall voltage originates from the misalignment of the current direction with the crystal axes and the intrinsic resistance anisotropy of $T_d$-$WTe_2$ [14]. In contrast, the third-harmonic voltages scale cubically with $I_\omega$ [Fig. 2(f)], clearly indicating that the third-order quantum nonlinearity of our device persists at room temperature. Additionally, we measure the second-harmonic voltages for comparison (Fig. S5 in Supplemental Material, Note 4 [40]), which are an order of magnitude smaller than $V^{3\omega}$, demonstrating that the third-order nonlinearity is dominant.

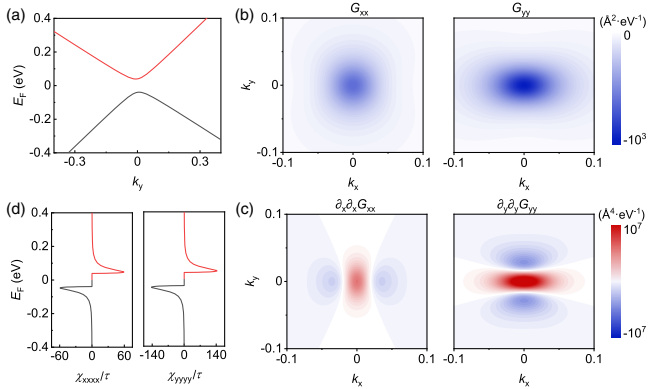

FIG. 1. Illustration of the QMQ in a tilted 2D massive Dirac model. (a) Band structure of the tilted 2D massive Dirac model as a function of $k_y$ at $k_x = 0$. (b) Distribution of the quantum metric components $G_{xx}$ (left) and $G_{yy}$ (right). (c) Distribution of the QMQ components $\partial_x\partial_x G_{xx}$ (left) and $\partial_y\partial_y G_{yy}$ (right), which contribute to the longitudinal third-order conductivities. (d) Calculated third-order conductivities $\chi_{xxxx}$ (left) and $\chi_{yyyy}$ (right) along the $x$ and $y$ directions as a function of Fermi energy. The unit of $k_x$ and $k_y$ is Å$^{-1}$, and the unit of $\chi_{xxxx}/\tau$ and $\chi_{yyyy}/\tau$ is $(e^4/\hbar^2)$ Å$^2 \cdot (eV)^{-1}$.

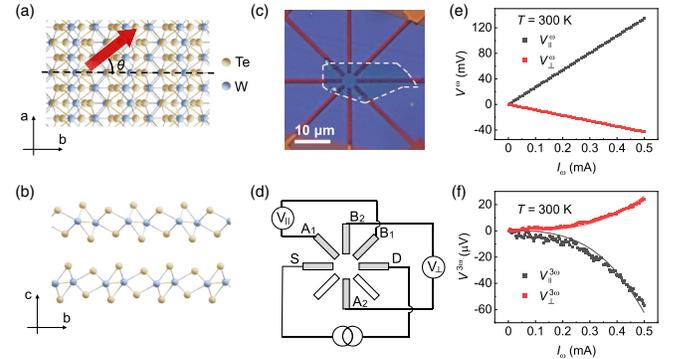

FIG. 2. Device measurements of $T_d$-$WTe_2$. (a), (b) Crystal structure in top (a) and side (b) views. Current is applied along angle $\theta$ relative to the $b$ axis, denoted by the red arrow in (a). (c) Optical image of a 6L device with eight electrodes. The white dashed line marks the border of the $T_d$-$WTe_2$ flake. (d) Measurement configuration illustration. (e), (f) First- and third-harmonic voltages in the longitudinal and transverse directions as a function of the driving current, measured at $\theta = 74°$ at 300 K. The solid lines in (f) are cubic fits to the data.

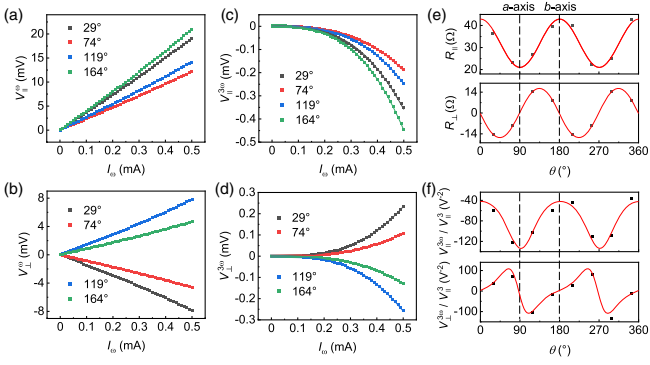
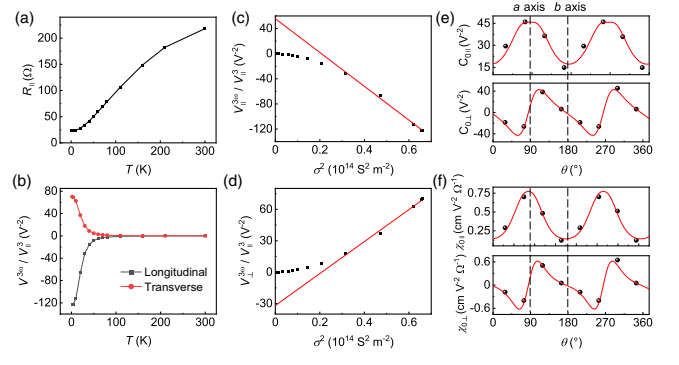

FIG. 3. Angle-resolved measurements of $T_d$-WTe$_2$ at 1.6 K. (a), (b) First-harmonic $I$-$V$ curves in the longitudinal and transverse directions at different angles, respectively. (c), (d) Third-harmonic $I$-$V$ curves in the longitudinal and transverse directions at different angles, respectively, with cubic fits shown as solid lines. (e) Longitudinal and transverse resistances as a function of $\theta$. (f) Third-order voltage ratios $V_{\parallel}^{3\omega}/V_{\parallel}^3$ and $V_{\perp}^{3\omega}/V_{\parallel}^3$ as a function of $\theta$. Red lines represent fitting results.

FIG. 4. Scaling analysis and angle-resolved measurements of the QMQ in $T_d$-WTe$_2$. (a), (b) Longitudinal resistance (a) and third-order voltage ratios (b) measured at $\theta = 74°$ as a function of temperature. (c), (d) $V_{\parallel}^{3\omega}/V_{\parallel}^3$ (c) and $V_{\perp}^{3\omega}/V_{\parallel}^3$ (d) as a function of $\sigma^2$. The scatter symbols represent experimental data with error bars smaller than the symbol size. Red lines indicate linear fits to the data below 30 K. (e) Fitting parameters $C_{0\parallel}$ and $C_{0\perp}$ as a function of $\theta$. (f) QMQ contributions $\chi_{0\parallel}$ and $\chi_{0\perp}$ as a function of $\theta$. Red lines indicate fitting results.

We conduct angle-resolved measurements at 1.6 K to verify the nonlinearity as an intrinsic property of the crystal. The angle $\theta$ is defined as the angle between the applied current and the crystal $b$ axis [Fig. 2(a)], determined by the polarized Raman spectrum [45] (Fig. S6 in Supplemental Material, Note 5 [40]). Longitudinal and transverse voltages are measured [Figs. 3(a) and 3(b)], from which we extract resistances $R_{\parallel}$ and $R_{\perp}$, respectively. A small deviation from Ohm's law due to third-order nonlinearity is observed, with further details provided in Supplemental Material, Note 6 [40]. A clear twofold angular dependence is observed [Fig. 3(e)], and the resistances are fitted by the formula derived from the crystal point group $Pm$: $R_{\parallel}(\theta) = R_b \cos^2\theta + R_a \sin^2\theta$ and $R_{\perp}(\theta) = (R_a - R_b)\sin\theta\cos\theta$, where $R_b$ and $R_a$ are the resistances along the crystal $b$ and $a$ axes, respectively. The resistance anisotropy $r = R_a/R_b$ is found to be 0.49 at 1.6 K.

Moreover, the third-harmonic voltages exhibit a cubic dependence on the driving current at all measured angles [Figs. 3(c) and 3(d)]. We extract the third-order voltage ratios $V^{3\omega}/V_{\parallel}^3$ and illustrate them in Fig. 3(f). Both the longitudinal and transverse $V^{3\omega}$ show angular dependence consistent with the crystal symmetry (see Supplemental Material, Methods [40]), revealing four independent third-order conductivity components: $\chi_{11} = \chi_{aaaa}$, $\chi_{22} = \chi_{bbbb}$, $\chi_{12} = \chi_{aabb}$, and $\chi_{21} = \chi_{bbaa}$. We fit the third-order voltage ratios as a function of $\theta$ and obtain the parameters $\chi_{11} = -0.3$ cm/(V$^2$ Ω), $\chi_{22} = -2.3$ cm/(V$^2$ Ω), $\chi_{12} = -0.8$ cm/(V$^2$ Ω), and $\chi_{21} = 0.1$ cm/(V$^2$ Ω). Notably, the measured $\chi_{22}$ is an order of magnitude larger than $\chi_{11}$, consistent with the tilted 2D massive Dirac model [Fig. 1(d)] and corresponding to the band anisotropy of WTe$_2$ [42].

Additionally, when the applied current is along the crystal $b$ or $a$ axes, the third-order Hall voltage vanishes while the longitudinal voltage remains finite [Fig. 3(f)]. This is due to the mirror symmetry that forbids the third-order Hall response for transport along these crystal axes. In contrast, the longitudinal third-order nonlinearity is subject to fewer symmetry constraints, offering broader potential applications.

To further investigate the origin of the observed third-order nonlinearity, we conduct experiments at temperatures ranging from 1.6 to 300 K to vary the transport scattering strength of WTe$_2$. As shown in Fig. 4(a), the longitudinal resistance measured at $\theta = 74°$ increases with temperature, indicating metallic behavior with a residual-resistance ratio of 9.4. In contrast, the third-order voltage ratios decrease monotonically, showing variations greater than 3500 times in magnitude [Fig. 4(b)]. This significant reduction highlights the sensitivity of the third-order nonlinearity to scattering strength. According to semiclassical transport theory [35], the intrinsic QMQ contribution is proportional to the scattering time ($\tau$), while extrinsic skew scattering contributions scale as $\tau^3$. Because of the time-reversal symmetry preserved in $T_d$-WTe$_2$, there are no $\tau^0$ and $\tau^2$ terms, as seen in the mechanisms of the anomalous Hall effect [3]. Thus, the scaling relation of the third-order nonlinearity is expressed as

$$\frac{V^{3\omega}}{V_{\parallel}^3} = C_1 \sigma^2 + C_0, \quad (2)$$

where $C_1$ and $C_0$ represent the skew scattering and QMQ contributions, respectively, and $\sigma$ is the longitudinal linear conductivity. Notably, the longitudinal nonlinearity follows the same scaling relation framework as the transverse nonlinearity, reflecting different perspectives of the quantum metric. We extract the third-order voltage ratios at different temperatures [Figs. 4(c) and 4(d)] and observe a good fit with Eq. (2) below 30 K. The deviation from the linear fitting under high temperatures may be attributed to the temperature driving Fermi surface shift [46–48], which significantly affects the transport effects due to the semimetallic property of $WTe_2$ (Supplemental Material, Note 8 [40]). The obtained QMQ contributions at low temperature are $C_{0\parallel} = 55.6$ V$^{-2}$ and $C_{0\perp} = -31.8$ V$^{-2}$, 3 orders of magnitude larger than those in bulk $MoTe_2$, $WTe_2$ [33], and $TaIrTe_4$ [39]. Moreover, by varying the measurement angles, the QMQ contributions as a function of $\theta$ are depicted in Fig. 4(e). The QMQ-induced third-order nonlinear conductivities, calculated as $\chi_0 = C_0 \xi \sigma$, where $\xi$ is the geometric parameter of the device, are shown in Fig. 4(f). All data from different angles fit well with the crystal anisotropy. Specifically, the longitudinal conductivity $\chi_{0\parallel}$ reaches its maximum and minimum along the $a$ and $b$ axes, respectively, consistent with the band anisotropy of $WTe_2$. The transverse conductivity $\chi_{0\perp}$ approximately vanishes along the $a$ and $b$ axes, constrained by the mirror symmetries. Thus, the high-order quantum metric components, manifested as the quantum metric quadrupoles, are clearly demonstrated through the angle-resolved third-order nonlinear measurements.

Various side effects, such as capacitive and thermal effects, can influence nonlinear transport measurements [15]. To test for these, we conducted third-harmonic measurements across a range of driving frequencies from 17.777 to 177.77 Hz (Fig. S4, Supplemental Material, Note 3 [40]) and observed no frequency dependence, ruling out capacitive coupling. Additionally, Joule self-heating could induce a third-harmonic voltage ($V_{\text{th}}^{3\omega}$) proportional to the temperature derivative of resistance (d$R$/d$T$) [49]. However, the observed temperature dependence of $V_\parallel^{3\omega}$ in Fig. 4(b) does not align with the d$R$/d$T$ trend in Fig. 4(a), thereby excluding thermal effects as a major contributor.

In conclusion, our study reveals that the quantum metric quadrupole induces a giant longitudinal third-order nonlinear electric response in few-layer $WTe_2$ at room temperature. Unlike transverse responses, this longitudinal third-order nonlinearity is less restricted by symmetry and can occur even in isotropic systems, thus broadening the scope for characterizing diverse materials. Our findings, which link the observed third-harmonic nonlinearity to the quantum metric quadrupole, underscore the role of quantum geometric tensor multipole moments in higher-order harmonic measurements, paving the way for exploring exotic quantum geometric physics and device applications.

*Acknowledgments*—This work was supported by the National Natural Science Foundation of China (Grants No. 62425401 and No. 62321004), and Innovation Program for Quantum Science and Technology (Grant No. 2021ZD0302403).

# Giant Third-Order Nonlinearity Induced by the Quantum Metric Quadrupole in Few-Layer WTe$_2$

Xing-Yu Liu[1], An-Qi Wang[1,*], Dong Li[1], Tong-Yang Zhao[1], Xin Liao[1] and Zhi-Min Liao[1,2,*]

[1] State Key Laboratory for Mesoscopic Physics and Frontiers Science Center for Nano-optoelectronics, School of Physics, Peking University, Beijing 100871, China.

[2] Hefei National Laboratory, Hefei 230088, China.

*E-mail: anqi0112@pku.edu.cn; liaozm@pku.edu.cn

Table of contents:

**Methods**

**Note 1. Semiclassical theory of the third-order nonlinearity**

**Note 2. Calculations for the tilted 2D massive Dirac model**

**Note 3. Effect of driving frequency**

**Note 4. Second-harmonic measurements**

**Note 5. Polarized Raman measurements**

**Note 6. Deviation from Ohm's law**

**Note 7. Data from another $T_d$-WTe$_2$ device**

**Note 8. Temperature variation induced Fermi level shift in WTe$_2$**

## Methods

### Device fabrication

Few-layer flakes of $WTe_2$ were mechanically exfoliated from a bulk $WTe_2$ crystal (HQ Graphene) and transferred onto $Si/SiO_2$ substrates. The dry transfer technique was used to fabricate the devices. We picked up the capping h-BN and $WTe_2$ thin flakes using polycarbonate (PC) film and landed them onto the prefabricated Ti/Au electrodes. The flakes were identified by optical contrast, and the thickness was measured by atomic force microscopy. The crystal orientation of the $WTe_2$ flakes was identified by polarized Raman spectroscopy (WITec alpha 300).

### Transport measurements

The devices were measured in an Oxford cryostat. The first-, second-, and third-harmonic voltages were measured using Stanford Research Systems SR830 and SR865A lock-in amplifiers. The driving current was utilized at a frequency of 17.777 Hz, unless otherwise specified.

### Angular dependence of the resistance and third-order nonlinearity

The crystal symmetry of few-layer $T_d$-$WTe_2$ belongs to *Pm* point group. Here we consider the transport in two dimensions to simplify. The anisotropic resistivity tensor of few-layer $T_d$-$WTe_2$ is expressed as $\rho = \begin{pmatrix} \rho_b & 0 \\ 0 & \rho_a \end{pmatrix}$. Under an applied current $\mathbf{j} = j \begin{pmatrix} \cos\theta \\ \sin\theta \end{pmatrix}$, the in-plane electric field $\mathbf{E} = \rho \mathbf{j} = j \begin{pmatrix} \rho_b \cos\theta \\ \rho_a \sin\theta \end{pmatrix}$ is generated. Therefore, the longitudinal and transverse voltages are obtained as $V_\parallel = E_\parallel L_\parallel = jL_\parallel(\rho_b \cos^2\theta + \rho_a \sin^2\theta)$ and $V_\perp = E_\perp L_\perp = jL_\perp(\rho_a - \rho_b)\sin\theta\cos\theta$, where $L_\parallel$ and $L_\perp$ denote the electrode spacing along the current direction and perpendicular to the current direction, respectively. The resistances are expressed as $R_\parallel(\theta) = R_b \cos^2\theta + R_a \sin^2\theta$ and $R_\perp(\theta) = (R_a - R_b)\sin\theta\cos\theta$, respectively.

The third-order nonlinearity is characterized by the third-order conductivity $\chi^{(3)}$, given by $j_a^{(3)} = \chi_{abcd}^{(3)} E_b E_c E_d$. For the 2D case of the *Pm* point group, only four independent components exist, including $\chi_{11} = \chi_{xxxx}$, $\chi_{22} = \chi_{yyyy}$, $\chi_{12} = \chi_{xxyy} = \chi_{xyxy} = \chi_{xyyx}$, and $\chi_{21} = \chi_{yyxx} = \chi_{yxyx} = \chi_{yxxy}$. The coordinates x, y and z correspond to the crystal *b*-, *a*-, and *c*-axes, respectively. Therefore, under the in-plane electric field **E**, the generated third-order current $\mathbf{j}^{(3)}$ induces a third-order electric field $\mathbf{E}^{(3)} = \rho \mathbf{j}^{(3)} = j^3 \begin{pmatrix} \chi_{11}\rho_b^4 \cos^3\theta + 3\chi_{12}\rho_b^2\rho_a^2 \cos\theta \sin^2\theta \\ \chi_{22}\rho_a^4 \sin^3\theta + 3\chi_{21}\rho_b^2\rho_a^2 \cos^2\theta \sin\theta \end{pmatrix}$. The longitudinal and transverse components are obtained as $E_\parallel^{(3)} = (\cos\theta, \sin\theta)\mathbf{E}^{(3)}$ and $E_\perp^{(3)} = (-\sin\theta, \cos\theta)\mathbf{E}^{(3)}$. The voltage ratios $V^{(3)}/V_\parallel^3$ are expressed as

$$\frac{V_\parallel^{(3)}}{V_\parallel^3} = \frac{E_\parallel^{(3)}}{E_\parallel^3 L_\parallel^2} = \frac{1}{L_\parallel^2} \frac{\chi_{11}\rho_b^4 \cos^4\theta + \chi_{22}\rho_a^4 \sin^4\theta + 3(\chi_{12}+\chi_{21})\rho_b^2\rho_a^2 \cos^2\theta \sin^2\theta}{(\rho_b \cos^2\theta + \rho_a \sin^2\theta)^3}, \quad (S1)$$

$$\frac{V_\perp^{(3)}}{V_\parallel^3} = \frac{E_\perp^{(3)} L_\perp}{E_\parallel^3 L_\parallel^3} = \frac{L_\perp}{L_\parallel^3} \frac{\left(3\chi_{21}\rho_b^2\rho_a^2 - \chi_{11}\rho_b^4\right)\cos^3\theta \sin\theta + \left(\chi_{22}\rho_a^4 - 3\chi_{12}\rho_b^2\rho_a^2\right)\sin^3\theta \cos\theta}{(\rho_b \cos^2\theta + \rho_a \sin^2\theta)^3}. \quad (S2)$$

Through fitting the third-order voltages with Equations (S1) and (S2), the four third-order conductivity components can be determined. Moreover, when *θ* = 0° and 90°, the longitudinal voltage ratio is equal to $\chi_{11}\rho_b/L_\parallel^2$ and $\chi_{22}\rho_a/L_\parallel^2$, respectively, but the transverse one vanishes.

Further, in our scaling analysis, the obtained fitting parameter C₀ obeys the same framework as the voltage ratio. Therefore, by fitting the results with Equations (S1) and (S2), we can extract the QMQ-induced third-order conductivities, $\chi_{11}^0$, $\chi_{22}^0$, $\chi_{12}^0$, and $\chi_{21}^0$. The longitudinal and transverse third-order conductivities $\chi_\parallel$ and $\chi_\perp$ are calculated as $\chi_\parallel \approx C_0 L_\parallel^2 \sigma$ and $\chi_\perp \approx C_0 L_\parallel^3 \sigma/L_\perp$, respectively. Therefore, the fitting formula are expressed as

$$\chi_\parallel = \frac{\chi_{11}^0 \rho_b^4 \cos^4\theta + \chi_{22}^0 \rho_a^4 \sin^4\theta + 3(\chi_{12}^0 + \chi_{21}^0)\rho_b^2\rho_a^2 \cos^2\theta \sin^2\theta}{(\rho_b \cos^2\theta + \rho_a \sin^2\theta)^4}, \quad (S3)$$

$$\chi_\perp = \frac{\left(3\chi_{21}^0 \rho_b^2\rho_a^2 - \chi_{11}^0 \rho_b^4\right)\cos^3\theta \sin\theta + \left(\chi_{22}^0 \rho_a^4 - 3\chi_{12}^0 \rho_b^2\rho_a^2\right)\sin^3\theta \cos\theta}{(\rho_b \cos^2\theta + \rho_a \sin^2\theta)^4}, \quad (S4)$$

where $\sigma \approx \frac{1}{\rho_\parallel} = 1/(\rho_b \cos^2\theta + \rho_a \sin^2\theta)$ for estimation. Besides, it doesn't affect the results along the *b*- and *a*-axes as $\chi_\parallel = \chi_{11}^0$ at *θ* = 0° and $\chi_\parallel = \chi_{22}^0$ at *θ* = 90°, respectively.

**Note 1. Semiclassical theory of the third-order nonlinearity**

The third-order nonlinearity is derived from the semiclassical theory with second-order accuracy proposed and developed by Gao *et. al.* (ref. 36) and Liu *et. al.* (ref. 35). The applied electric field can induce a first-order correction to the Berry connection and Berry curvature, expressed as $\mathcal{A}_a^{(1)} = eG_{ab}(\mathbf{k})E_b$ and $\mathbf{\Omega}^{(1)} = \nabla_{\mathbf{k}} \times \mathcal{A}^{(1)}$, where $G_{ab}(\mathbf{k})$ is the Berry connection polarizability tensor (BCPT), given by

$$G_{ab}^n(\mathbf{k}) = 2\sum_{m \neq n} \frac{g_{ab}^{nm}}{\epsilon_n - \epsilon_m}. \tag{S5}$$

Here, $g_{ab}^{nm} = \text{Re}[\langle \partial_{k_a} u_n | u_m \rangle \langle u_m | \partial_{k_b} u_n \rangle]$ is the interband quantum metric. The quantum metric of the *n*th band is expressed as $g_{ab}^n = \sum_{m \neq n} g_{ab}^{nm}$. For a two-band model, the BCPT is proportional to the quantum metric, given by $G_{ab}^n(\mathbf{k}) = 2g_{ab}^n/(\epsilon_n - \epsilon_{n'})$, where $n'$ denotes the nearest band of the *n*th band. For multiband cases, the nearest band offers the largest contribution to BCPT. Therefore, the quantum metric is considered as a dominant contribution in BCPT-induced effects.

The correction of Berry curvature can generate a Hall current perpendicular to the electric field. Further, the correction of band energy cannot be ignored, expressed as

$$\epsilon^{(2)} = -\frac{e^2}{2}E_a G_{ab}(\mathbf{k})E_b. \tag{S6}$$

Therefore, the applied electric field can affect the band velocity as shown in Equation (1). Utilizing the distribution function *f* solved from the Boltzmann equation, the generated current is expressed as

$$\mathbf{j} = -e\int[d\mathbf{k}]\dot{\mathbf{r}}f = -e\int[d\mathbf{k}]\left(\tilde{v}_n + \frac{e}{\hbar}\mathbf{E} \times (\mathbf{\Omega} + \mathbf{\Omega}^{(1)})\right)f. \tag{S7}$$

We can obtain the third-order conductivity with terms proportional to $\tau^0, \tau^1, \tau^2, \tau^3$ and so on. The $\tau^1$ term is the intrinsic QMQ contribution as we discussed, and it is expressed as

$$\chi_{abcd}^{(3)} = \tau \frac{e^4}{\hbar^2}\int[d\mathbf{k}]\left\{(-\partial_{k_a}\partial_{k_b}G_{cd} + \partial_{k_a}\partial_{k_d}G_{bc} - \partial_{k_b}\partial_{k_d}G_{ac})f_0 + \frac{\hbar^2}{2}v_a v_b G_{cd} f_0''\right\}. \tag{S8}$$

For the longitudinal conductivities along the principal axes, we set the labels to the same and obtain $\chi^{(3)}_{aaaa} = \tau \frac{e^4}{\hbar^2} \int [dk](-\partial_{k_a}\partial_{k_a} G_{aa} f_0 + \frac{\hbar^2}{2} v_a^2 G_{aa} f_0'')$, which contains the QMQ component $\partial_{k_a}\partial_{k_a} G_{aa}$. This gives the calculated results as shown in Fig. 1.

Further, for better understanding of the properties of quantum metric and QMQ, we demonstrate a clear physical picture (ref. 26) in Fig. S1. The quantum metric describes the amplitude distance between two neighboring quantum states. Specifically, in Bloch bands (Fig. S1), the nearest states are the two states with a small shift in **k**, such as $|u_\mathbf{k}\rangle$ and $|u_{\mathbf{k}+d\mathbf{k}}\rangle$. If these two states originate from the same orbital (or spin) state, their separation is expected to be negligible, implying a zero-quantum metric. However, if there is band mixing, $|u_\mathbf{k}\rangle$ and $|u_{\mathbf{k}+d\mathbf{k}}\rangle$ may come from two different orbital states, resulting in a larger quantum metric. The QMQ, which quantifies variations in the quantum metric, is thus anticipated to be observable. In a word, systems with band mixing or band crossings can exhibit a large quantum metric, leading to substantial QMQ-induced third-order nonlinearities.

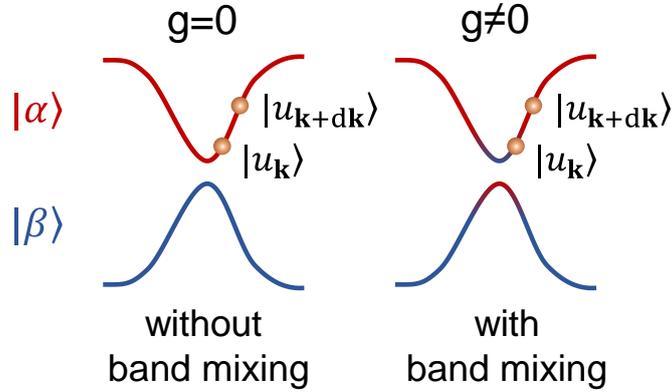

**Fig. S1: Illustration of the quantum metric in different band structures.** The quantum metric can possess a significant value in a band structure with band mixing or band crossing.

We also made comprehensive analysis of different origins of third-order nonlinearity in two-dimensional systems, which is summarized in Table. R1. To extract the QMQ contribution, both the symmetry analysis and scaling analysis are necessary.

**Table S1: Comparison of different mechanisms of third-order nonlinear transport effects.** *P* and *T* represent the spatial inversion symmetry and time-reversal symmetry. Mirror symmetry here is only considered as the case where the mirror plane is perpendicular to the sample plane. Allowed: ✔ ; Forbidden: ⊘.

| Property | *P* symmetry | *T* symmetry | Mirror symmetry with *T* | Isotropy with *T* | τ |
|---|---|---|---|---|---|
| **Berry Curvature Quadrupole** | ✔ (only Hall) | ⊘ | ⊘ | ⊘ | $\tau^2$ |
| **Quantum Metric Quadrupole** | ✔ | ✔ | ✔ (only longitudinal along axes) | ✔ (only longitudinal) | $\tau^1$ |
| **Nonlinear Drude Conductivity** | ✔ | ✔ | ✔ (only longitudinal along axes) | ✔ (only longitudinal) | $\tau^3$ |
| **Additional Scattering Mechanisms** | ✔ | ✔ | ✔ (only longitudinal along axes) | ✔ (only longitudinal) | Mainly $\tau^3$ |

**Note 2. Calculations for the tilted 2D massive Dirac model**

The tilted 2D massive Dirac model we consider is given by

$$H_d(k) = tk_y + v_x k_x \sigma_x + v_y k_y \sigma_y + \Delta \sigma_z, \quad (S9)$$

where $\sigma_x$, $\sigma_y$, and $\sigma_z$ are Pauli matrices. Here, $t$ represents the band tilting along the $y$-direction, and $v_x$, $v_y$, and $\Delta$ are model parameters. We set $v_y = 1$ eV·Å, $v_x = 0.8\ v_y$, $t = 0.2\ v_y$, and $\Delta = 0.04$ eV, with results shown in Fig. 1.

For isotropic band structures, we take $v_x = v_y = 1$ eV·Å, $t = 0$, and $\Delta = 0.04$ eV for Fig. S2 and $\Delta = 0$ eV for Fig. S3. Similar results can be observed, where the third-order longitudinal conductivities reach maxima near the band edge and survive even as the band gap vanishes. Besides, owing to the rotation symmetry of the band structure, the transverse third-order conductivities vanish, suggesting the longitudinal third-order nonlinearity as an efficient probe in a wider range of materials.

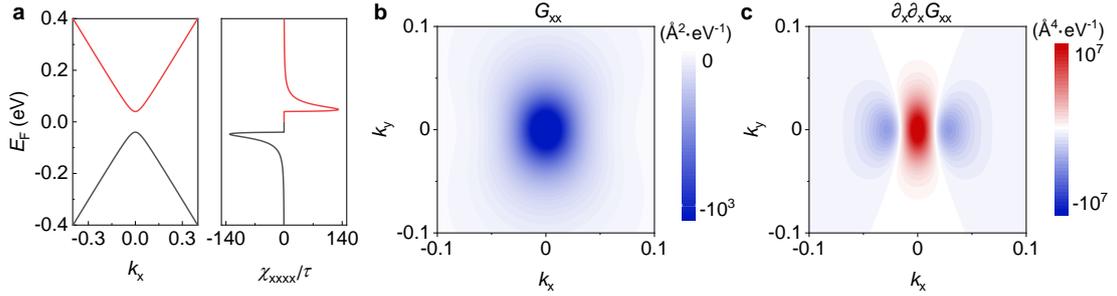

**Fig. S2: Third-order nonlinearity in a 2D massive Dirac model. a**, Band structure of the isotropic 2D massive Dirac model, showing the third-order conductivity $\chi_{xxxx}$ as a function of Fermi energy. **b,c**, Distribution of quantum metric $G_{xx}$ and its second derivative $\partial_x \partial_x G_{xx}$ (QMQ). Units: $k_x$ and $k_y$ in Å$^{-1}$; $\chi_{xxxx}/\tau$ in $\frac{e^4}{\hbar^2}$Å$^2 \cdot$(eV)$^{-1}$.

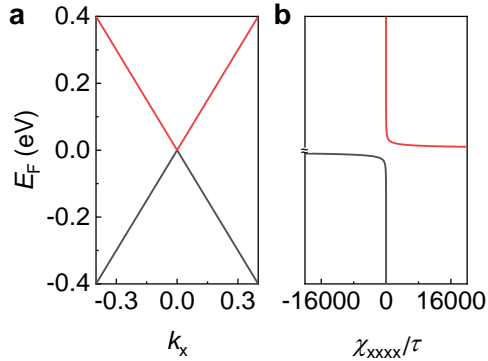

**Fig. S3: Calculated results of third-order nonlinearity in a 2D Dirac model.** The unit of $k_x$ is Å$^{-1}$, and the unit of $\chi_{xxxx}/\tau$ is $\frac{e^4}{\hbar^2}$Å$^2 \cdot$(eV)$^{-1}$.

## Note 3. Effect of driving frequency

We vary the frequency of the applied current $I_\omega$ and measure the third-order nonlinear response. As shown in Fig. S4, no frequency dependence is observed ranging from 17.777 Hz to 177.77 Hz, excluding the spurious capacitive coupling effect.

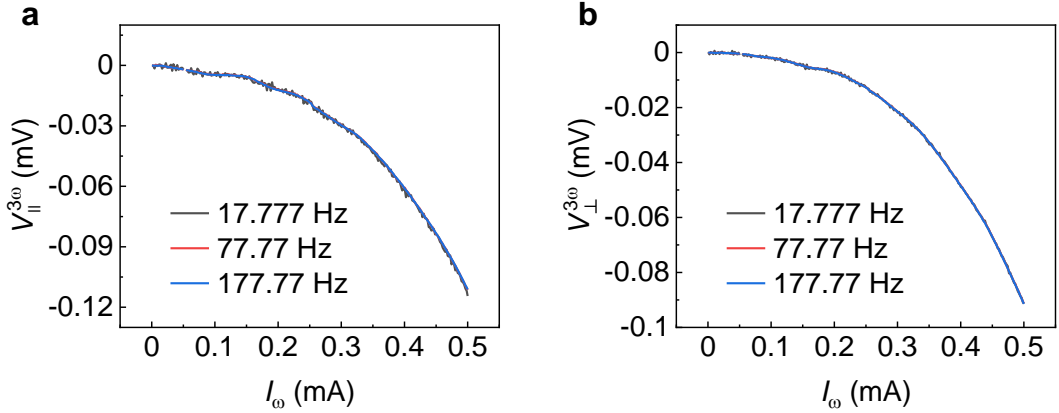

**Fig. S4: Third-order longitudinal and transverse voltages as a function of $I_\omega$ for driving frequencies ranging from 17.777 Hz to 177.77 Hz, measured at $\theta = 299°$ and 110 K.**

## Note 4. Second-harmonic measurements

Second-harmonic voltages were measured at 300 K for comparison. As shown in Fig. S5, the second-order signals are at the microvolt level with significant noise fluctuations, an order of magnitude smaller than the third-order voltages (Fig. 2f), confirming third-order nonlinearity as the dominant contribution.

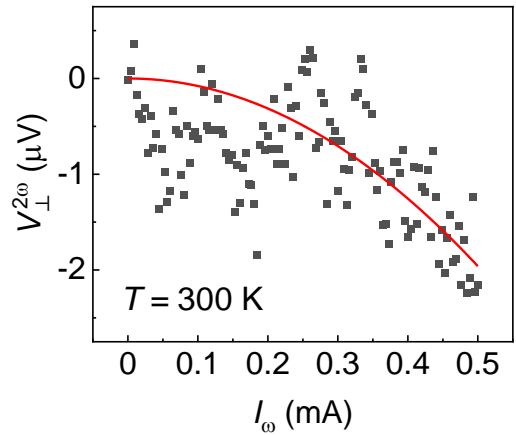

**Fig. S5: Second-order transverse voltages as a function of the driving current, measured at $\theta = 74°$ and 300 K.**

**Note 5. Polarized Raman measurements**

Figure S6b shows angle-resolved polarized Raman spectra of our $T_d$-WTe$_2$ device in the parallel configuration (ref. 44). The relative intensity of the Raman mode A$_g$ (A$_1$) at ~212 cm$^{-1}$ is used to determine the crystal axes (Fig. S6c).

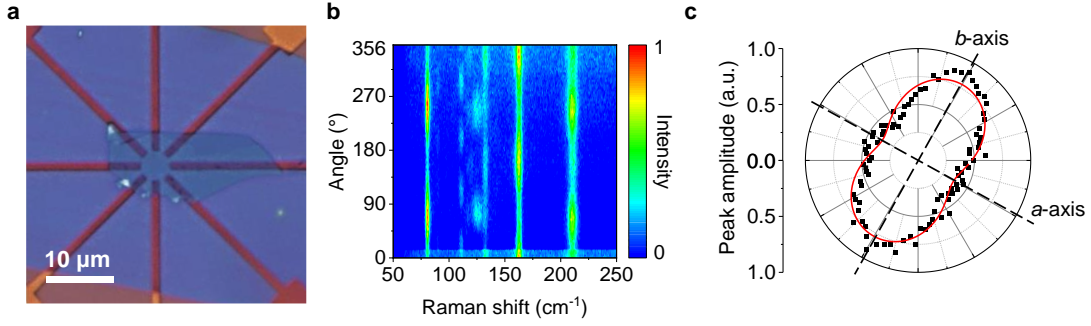

**Fig. S6: a,** Optical image of a 6L $T_d$-WTe$_2$ device. **b,** A typical angle-resolved polarized Raman spectra of $T_d$-WTe$_2$ in the parallel configuration. **c,** The angle-dependent relative intensity of the A$_g$ (A$_1$) mode at ~212 cm$^{-1}$.

**Note 6. Deviation from Ohm's law**

We observe deviations from Ohm's law in the first-harmonic current-voltage characteristics, as shown in Fig. S7. At 1.6 K, when the applied current exceeds 0.3 mA, the first-harmonic longitudinal and transverse voltages deviate from the linear fits (red lines in the upper panels of Figs. S7a and S7b). These deviations, extracted and shown to fit a cubic dependence, are attributed to third-order nonlinearity in the device.

The total voltage under an applied AC current $I_\omega$ is given by $V = V^{(1)} + V^{(3)} = RI_\omega + R^{(3)}I_\omega^3$, where $R$ is the linear resistance along the measurement direction, and $R^{(3)}$ is the third-order resistance. For $I_\omega = I \sin \omega t$, this becomes

$$V = RI \sin \omega t + R^{(3)}(I \sin \omega t)^3 = \left(RI + \frac{3}{4}R^{(3)}I^3\right)\sin \omega t - \frac{1}{4}R^{(3)}I^3 \sin 3\omega t.$$

In our lock-in measurement, the observed first-harmonic voltage is the first term, $V^\omega = RI + \frac{3}{4}R^{(3)}I^3$, and the third-harmonic voltage is $V^{3\omega} = -\frac{1}{4}R^{(3)}I^3$. This cubic dependence in $V^\omega$ confirms third-order nonlinearity in the device.

Further, we examine the relation between the cubic term in $V^{\omega}$ and $V^{3\omega}$. The cubic term in $V^{\omega}$ is proportional to $V^{3\omega}$ with a ratio of -3, consistent with expectations, as shown in Figs. S7c and S7d for both longitudinal and transverse voltages.

In analyzing our results, we account for this third-order contribution, particularly in resistance fitting. At high temperatures, third-order effects are negligible, resulting in a linear dependence of $V^{\omega}$ on the current. However, at low temperatures, significant cubic deviations require incorporating this term to accurately determine device resistance.

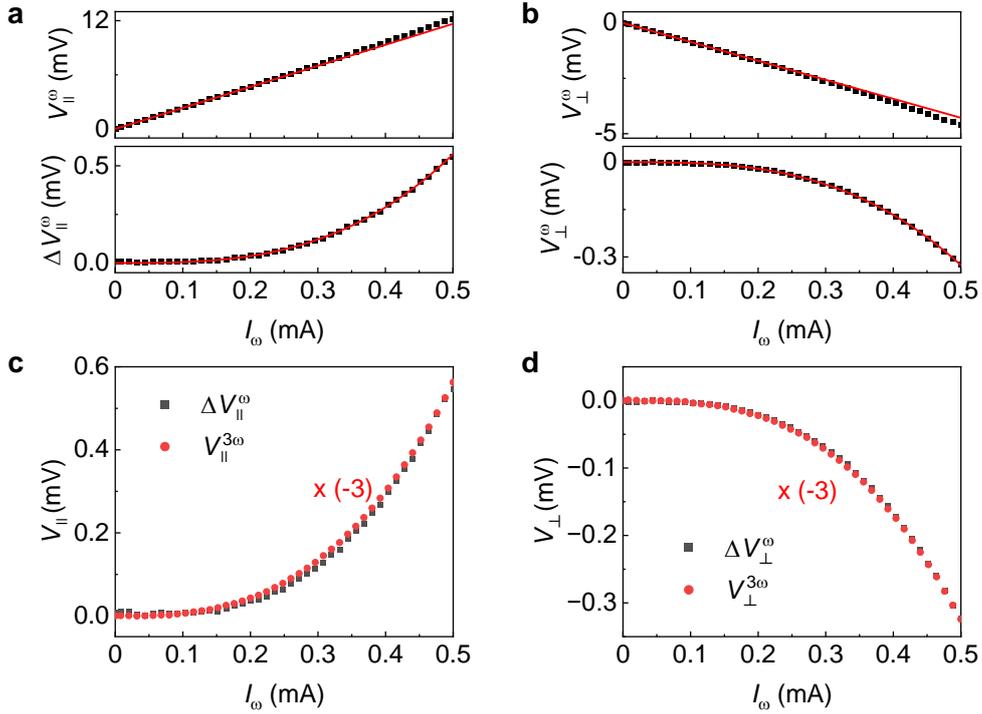

**Fig. S7: Deviation from Ohm's Law at 1.6 K. a,** First-harmonic voltages measured in the longitudinal direction at 1.6 K and $\theta = 74°$. Red line (linear fit) in the upper panel highlights a clear deviation, shown with cubic fitting in the lower panel. **b,** Transverse measurements showing similar results as in (**a**). **c,** Comparison of the nonlinear component in $V^{\omega}$, that is $\Delta V^{\omega}$, with the measured $V^{3\omega}$ in the longitudinal direction, where $V^{3\omega}$ is scaled by -3. **d,** Similar comparison as in (**c**) for the transverse direction.

## Note 7. Data from another $T_d$-WTe$_2$ device

We fabricate another 10L $T_d$-WTe$_2$ device and conduct the similar third-order nonlinearity measurements at 1.6 K. As shown in Fig. S8, the first-harmonic longitudinal and transverse voltages show a linear dependence on the driving current. The third-harmonic longitudinal and transverse voltages exhibit a clear cubic dependence. Compared to the results in Fig. 3, the resistance of this device is smaller than the 6L device, and the third-order voltages are approximately one order of magnitude smaller. The results measured from this 10L device are consistent with the reported studies (ref. 38).

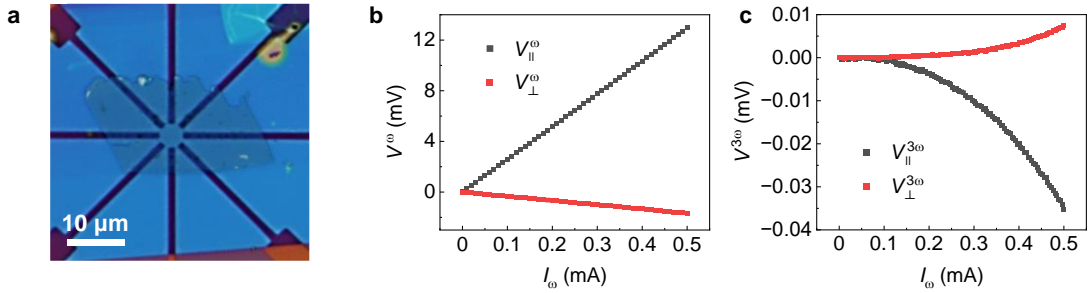

**Fig. S8: a,** Optical image of a 10L $T_d$-WTe$_2$ device. **b,** The first-harmonic longitudinal and transverse voltages as a function of $I_\omega$ measured at 1.6 K. **c,** The third-harmonic longitudinal and transverse voltages as a function of the driving current at 1.6 K.

## Note 8. Temperature variation induced Fermi level shift in WTe$_2$

WTe$_2$ shows a pronounced Fermi level shift with temperature, as demonstrated by previous studies (refs. 45-47), including ARPES measurements showing a ~50 meV shift over 120 K. Our model calculations in Fig. 1d indicate that third-order conductivities are observable only within ~100 meV of the band edge, suggesting that the QMQ contribution may decrease as the Fermi level shifts with increasing temperature.

To verify this, we performed magnetoresistance and Hall resistance measurements on a 7-layer (7L) WTe$_2$ device, previously studied in ref. 49. Hall resistance decreases with increasing temperature, and a two-carrier model fit reveals an increase in electron density and decrease in hole density as temperature rises. For accurate QMQ extraction, we selected data below 30 K, where carrier density variation is minimal, and scattering time dominantly affects conductivity. Fitting within this range (Fig. 4 and Fig. S9) allowed us to isolate Drude and QMQ contributions at low temperatures.

At higher temperatures, electron density increases while hole density becomes negligible, indicating a Fermi level shift away from the charge compensation point, where QMQ is smaller. Consequently, QMQ contributions diminish as temperature increases. Furthermore, QMQ effects and other scattering mechanisms both depend on scattering time: the QMQ contribution scales with $\tau^1$, while Drude and additional scattering contributions scale mainly with $\tau^3$. At higher temperatures, dynamic scattering processes, like phonon scattering, are greatly enhanced and $\tau$ is suppressed, resulting in increased resistance and reduced third-order nonlinear effects, as shown in Figs. 4a and 4b. The similar behavior of third-order effects is also observed in previous studies on 8-10L $WTe_2$ (ref. 38).

Additionally, as temperature increases, $WTe_2$ resistance rises, potentially amplifying thermal effects. The thermal third-order nonlinear voltage $V_{th}$ scales with both the resistance $R$ and its temperature derivative $dR/dT$, both of which increase with temperature. This thermal component may eventually dominate the third-order nonlinearity over QMQ and other scattering mechanisms.

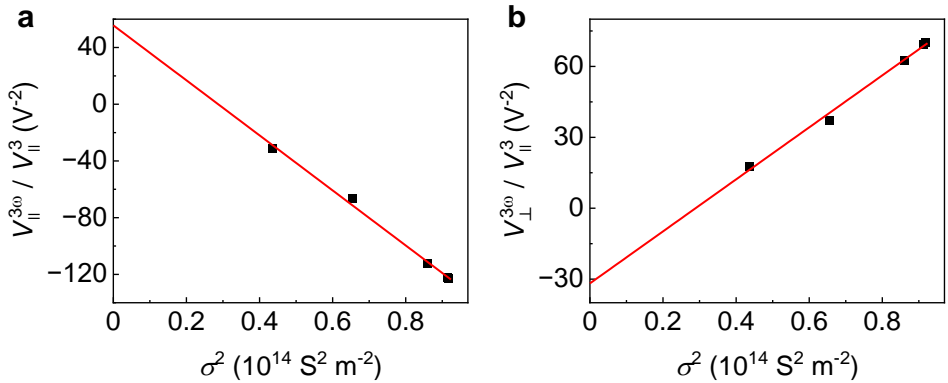

**Fig. S9: Scaling fitting using the data measured below 30 K. a,** $V_\parallel^{3\omega}/V_\parallel^3$ and **b,** $V_\perp^{3\omega}/V_\parallel^3$ as a function of $\sigma^2$.